\documentclass[letterpaper,twocolumn,english,showpacs,prb,aps]{revtex4}
\usepackage[T1]{fontenc}
\usepackage[latin1]{inputenc}
\usepackage{graphicx}
\usepackage{amssymb}

\makeatletter

\large  

\input epsf

\usepackage{babel}
\makeatother
\begin{document}

\title{Ground-state properties of a large 
Coulomb-blockaded quantum dot}

\author{Eugene B. Kolomeisky $^{(1)}$, Michael Timmins $^{(1)}$ and 
Ryan M. Kalas $^{(2)}$}

\affiliation{$^{(1)}$ Department of Physics, University of Virginia, P. O. Box 400714,
Charlottesville, Virginia 22904-4714, USA\\
$^{(2)}$  Department of Physics and Astronomy, Washington State
University, Pullmann, Washington, 99164-2814, USA}

\begin{abstract}
Using renormalization-group techniques we analyze equilibrium properties
of a large gated quantum dot coupled via a long and narrow channel to a reservoir of electrons.  Treating the electrons
in the channel as one-dimensional and interacting, we demonstrate that for
nearly-open dot and not very strong spin fluctuations the
ground-state properties as a function of the gate voltage are 
non-analytic at the points of half-integer average dot population.  
Specifically, the exact result of K. A. Matveev, Phys. Rev. B
\textbf{51}, 1743 (1995), that the dot capacitance shows periodic
logarithmic singularities is rederived as a special case corresponding
to non-interacting electrons.  We demonstrate that this conclusion also
holds in the presence of $SU(2)$ spin symmetry, and argue that
logarithmic singularities persist as long as the Coulomb blockade is
destroyed which will be the case for sufficiently large tunneling and
not very strong interparticle repulsions.  We
show that interparticle repulsions aid the Coulomb blockade to survive
disordering effect of zero-point motion provided the tunneling is
sufficiently weak.   Upon increase of tunneling, the Coulomb blockade
disappears through a Kosterlitz-Thouless phase transition
characterized by a
dot population jump whose magnitude is only determined by interparticle
repulsions in the channel.   Similar conclusions are applicable
to systems without $SU(2)$ spin symmetry except that
the logarithmic singularity of the capacitance is replaced by a
power-law non-analyticity whose exponent characterizes the degree of
spin fluctuations.

\end{abstract}


\maketitle

\section{Introduction}

Quantum dots are mesoscopic objects whose properties, like those
of ordinary atoms, can vary appreciably upon addition or removal of a
single electron \cite{Aleiner}.  This fact makes them an active
playground of studying the many-body physics under the
conditions of confinement.

A simple picture of a dot pertinent to understanding its
equilibrium properties
is that of a container of electrons connected to a
lead via a single constriction and placed nearby a gate
electrode.  The role of the latter is to tune the chemical
potential of the dot to host a desirable number of
electrons which is generally not integer.  On the other hand, the
electrons are discrete, and can only hop onto or off the dot one at a
time.  Therefore they
enter or leave the dot only at special values of the gate
voltage corresponding to half-integer dot population.  This effect
well identified by the early 1990's is called
the Coulomb blockade \cite{Likharev}.   One of its hallmarks is
the dependence of the population of the dot on the gate voltage
consisting of plateaus corresponding to an integer number of electrons
on the dot - the famed Coulomb staircase.

Quantum-mechanical tunneling of charge between the connecting lead and the dot
has been argued to smear the classical Coulomb staircase even at zero
temperature \cite{glaz}.  This idea is supported by a connection of the
quantum dot problem to a Kondo problem
\cite{Matveev91}.  In the limit of almost zero tunneling the population
dynamics on the dot can be approximated
by that of a two-level system, with the levels corresponding to
population states
of the dot differing by unity \cite{Matveev91}.  Changing the
population on the dot is then akin to flipping an impurity spin-1/2 while
the deviation of the average dot population from the half-integer
value is analogous to a magnetic field applied to the Kondo spin.  For a
single-channel Kondo problem pertinent to the spinless version of the
quantum dot problem, the spin susceptibility is
finite at zero applied field.  This implies smearing
of the population jumps with variation of the gate voltage.

Complementary treatments of the strong-tunneling regime
\cite{Flensberg,Matveev95} led to a conjecture that the rounded Coulomb staircase of the weak-tunneling limit
smoothly evolves as tunneling is increased into a strictly linear function of
the gate voltage.

A spinless version of the Coulomb blockade problem has been recently
revisited \cite{KKQ} where additionally it was assumed that the dot is
connected to a one-dimensional reservoir so that interparticle
interactions have to be taken into account.  It
has been demonstrated that as the strength of the tunneling is tuned,
changing the dot from open to closed, a modified Coulomb staircase
reappears under certain circumstances.  This was arrived at by
mapping the zero-temperature dynamics of interacting electrons to
the classical statistical mechanics of a one-dimensional Ising model with
ferromagnetic interactions decaying as the inverse-square of distance.
This led to the conclusion that all experimental systems must be
either of the Kondo subclass  (to which the connection to the Kondo problem
\cite{Matveev91} applies) or of a novel tricritical type.

Here we reconsider the problem of the Coulomb-blockaded quantum dot and
take into account both the electron spin and electron-electron
interactions in the channel connecting the dot with the electron reservoir.  This is relevant because
the strongest argument in favor of the Kondo analogy has been
given for the non-interacting version of this problem:  Matveev
\cite{Matveev95} found exactly that in the high-transparency limit, the capacitance of the dot has a
logarithmic singularity whenever the average population of the dot is
half-integer.  The weak-tunneling analysis of this problem
\cite{Matveev91} which used the solution of the {\it two-channel} Kondo
problem also found a logarithmic divergence which led to
the contention that this singularity of the capacitance
persists for arbitrary tunneling.   These predictions \cite{Matveev95}
found partial experimental support \cite{Berman} thus making the issue practically relevant.

The present work investigates the problem in a phenomenological
fashion by using renormalization-group techniques which have an
advantage of 	generality, thus complementing earlier
calculations.  Specifically, we show that the
logarithmic singularity of the capacitance \cite{Matveev91,Matveev95}
is a consequence of the assumption that the electrons in the
constriction are non-interacting.  We also point out that the same
conclusion holds in the presence of $SU(2)$ spin symmetry as long as
the Coulomb blockade is destroyed.  Moreover,
renormalization-group methods allow us to see clearly possible
limitations of existing approximations, thus motivating future work.

The organization of this paper is as follows.  In Section II we set up
the problem in the form of a zero-dimensional quantum field theory describing
the dynamics of the charge and spin of the dot. For a nearly-open dot
(Section III), the population of the dot tracks the gate
voltage and only the spin dynamics described by a sine-Gordon like
theory is relevant.  In Section IIIA this
field theory is studied perturbatively in small reflection
amplitude.  This exercise demonstrates that physically
interesting cases are beyond the perturbative approach.  Section
IIIB contains a heuristic argument which allows us to guess the answers
for situations when the perturbation theory fails.  These conclusions
are placed on a sounder footing in Section IIIC where
renormalization-group analysis is employed.  The novel feature
here is a systematic calculation of the renormalization of the
ground-state energy which allows, for example, to exactly reproduce
the logarithmic divergence of the capacitance \cite{Matveev95}.  In
Section IIID we draw a parallel between these results and those found
earlier for the standard sine-Gordon theory.

In Section IV we study the opposite limit of  a nearly-closed dot.  For
half-integer average dot population the dynamics of electrons hopping onto and
off the dot is approximated by that of a two-state system.  The
zero-temperature properties of the latter can be inferred from
the classical statistical mechanics of a two-component model resembling
the one-dimensional Ising model with interactions decaying as
the inverse-square of distance.  An ``entropy'' argument which in
Section IVA is applied to this model provides us with the range
of stability of the regime of zero tunneling.  These conclusions are
refined in Section IVB via a renormalization-group calculation
perturbative in tunneling amplitude.  This calculation indicates that
whenever the tunneling is relevant, the non-analytic behavior of the
ground-state properties of the dot has to be captured by the results
found for a nearly-open dot.  In Section IVC this statement is made more
quantitative via a heuristic argument which shows how the
weak-tunneling results away from half-integer dot population can be
inferred from those of Sections III and IVB.  We conclude (Section V)
by summarizing our results and emphasizing possible limitation of the two-state
approximation of the hopping dynamics for a nearly-closed dot.

\section{Statement of the problem}

Following Matveev \cite{Matveev95} we consider a large quantum dot of
capacitance $C_{0}$ whose charging energy $E_{C} = e^{2}/2C_{0}$ is
significantly larger than the distance between the energy
levels on the dot.  The dot is placed near a gate and connected to a
reservoir of spin-1/2 electrons through a channel which we assume to
be so narrow and long that the electrons in the channel are
essentially one-dimensional and their interactions have to be taken
into consideration.  Although there are no known experimental
realizations of this model, appropriate experiments might become
possible in the future.  Additionally, this represents a generalization
of the model due to Matveev \cite{Matveev95} in a way that adds
interactions violating $SU(2)$ spin symmetry because it is
interesting from theoretical point of view.  The corresponding long-wavelength low-energy Euclidian action
has the form
\begin{eqnarray}
\label{action}
A& =& {\frac{1}{2}}\int \sum\limits_{\nu = c,s}
dxd\tau\rho_{\nu} \left (({\frac{\partial u_{\nu}}{\partial \tau}})^{2} +
c_{\nu}^{2}({\frac{\partial u_{\nu}}{\partial x}})^{2}\right )\nonumber\\& +& \int d\tau V(u_{\uparrow}(0,\tau), u_{\downarrow}(0,\tau))
\end{eqnarray}
The first harmonic term describes interacting one-dimensional electrons
inside the channel.   Here the displacements of the electrons of each
spin population with respect to their classical positions are
represented by the fields $u_{\uparrow}(x,\tau)$ and $u_{\downarrow}(x,\tau)$
where arrows indicate the spin direction, $x$ is the one-dimensional
coordinate along the channel and $\tau$ is the imaginary time.
Written in terms of the center of mass (``charge'') and reduced relative
displacement (``spin'') fields
\begin{equation}
\label{chargespin}
u_{c} = {\frac {u_{\uparrow} +
u_{\downarrow}}{2}}, ~~~ u_{s} = {\frac {u_{\uparrow} -
u_{\downarrow}}{2}},
\end{equation}
correspondingly, the first term of (\ref{action}) is diagonal.  The
parameters $\rho_{\nu}$ and $c_{\nu}$ are the macroscopic mass densities and
sound velocities in the charge and spin sectors \cite{RMP}.

The harmonic (Luttinger) liquid exponents
\begin{equation}
\label{exponents}
g_{\nu} = {\frac {\pi\hbar(2n)^{2}}{\rho_{\nu}c_{\nu}}}
\end{equation}
characterize interparticle interactions in the charge ($\nu = c$) and
spin ($\nu = s$) channels \cite{RMP}.  They are formed out of the
parameters of the harmonic part of the action
(\ref{action}) and the one-dimensional electron density $n$ within
each spin population.  Although non-interacting
electrons correspond to $g_{c,s} = 2$, in what follows we will
consider more general values of the exponents.

We also note that either spin population creates a periodic potential
for its opposite spin counterpart which may cause spin ordering.  For
weak short-range interaction between the electrons of opposite spin this
can be modeled by adding to (\ref{action}) a term
proportional to $\int dxd\tau\cos4 \pi n u_{s}$.  We however assume that
such interactions are irrelevant in the renormalization-group sense \cite{Ma},
and thus the spin ordering does not take place. This would be the
case, for example, in the presence of $SU(2)$ spin symmetry (the
continuum Hubbard model with repulsive interactions can serve as an
example) where $g_{s} = 2$.  More generally we will require that the degree of
quantum fluctuations in the spin channel is sufficiently large \cite{RMP},
specifically $g_{s} \ge 2$.

The second term of the action (\ref{action}) is due to
the tunneling and Coulomb repulsion on the dot and the arguments of the
potential $V$ are evaluated  at $x = 0$ which can be taken to be
the narrowmost place of the channel \cite{Matveev95}.

Since the interactions in (\ref{action}) are restricted to a
point, the harmonic degrees of freedom away from $x = 0$ can be
integrated out.  This is equivalent to minimizing the action
(\ref{action}) with fixed values of the fields $u_{c,s}$ at the
origin followed by the substitution of the solution back into
(\ref{action})  and evaluation of the outcome.  The conditions
$\delta A/\delta u_{c,s}$ = 0 reduce to the Laplace equations
\begin{equation}
\label{Laplace}
{\frac {\partial^{2} u_{c,s}}{\partial \tau^{2}}} + c_{c,s}^{2}{\frac
{\partial^{2} u_{c,s}}{\partial x^{2}}} =0
\end{equation}
whose solutions satisfying the boundary conditions
$u_{c,s}(0,\tau) \equiv u_{c,s}(\tau)$
have the form
\begin{equation}
\label{solution}
u_{c,s}(x,\tau) =
\int\limits_{-\infty}^{+\infty}{\frac{d\omega}{2\pi}}u_{c,s}(\omega)
\exp(i\omega \tau - |\omega| |x|/c_{c,s})
\end{equation}
where we introduced the Fourier transforms
$u_{c,s}(\omega) = \int d\tau u_{c,s}(\tau)
\exp(-i\omega \tau)$.  Substituting (\ref{solution}) in (\ref{action}),
performing the integration, and introducing two new fields $\theta_{c}$
and $\theta_{s}/2$ corresponding to the fluctuating electron number and spin
on the dot, respectively
\begin{equation}
\label{numberspin}
\theta_{c} = n(u_{\uparrow} + u_{\downarrow}) = 2nu_{c},~~~ \theta_{s} =
n(u_{\uparrow} - u_{\downarrow}) = 2nu_{s}
\end{equation}
we find the effective action
\begin{equation}
\label{effaction}
A_{eff} =  \pi \hbar \sum \limits_{\nu = c,s}
\int\limits_{|\omega| \le \Lambda_{0}}  {\frac
{d\omega}{2\pi g_{\nu}}}|\omega||\theta_{\nu}(\omega)|^{2} + \int d\tau V(\theta_{c}, \theta_{s})
\end{equation}
where $\Lambda_{0}$ is the high-frequency cutoff.  Here without the loss of generality we
restrict ourselves to the sharp cutoff model with all the
$|\omega| > \Lambda_{0}$ modes eliminated.

For the problem at hand the potential $V$ entering the actions
(\ref{action}) and (\ref{effaction}) is a sum of two contributions:
\begin{equation}
\label{effpotential}
V = E_{C}(\theta_{c} - N)^{2} + U(u_{\uparrow},u_{\downarrow}) \equiv
E_{C}(\theta_{c} - N)^{2} + U(\theta_{c}, \theta_{s})
\end{equation}
where the first term is the charging energy with dimensionless
parameter $N$ proportional to the gate voltage \cite{Matveev95}.
The second, tunneling term $U(u_{\uparrow},u_{\downarrow})$ is periodic
in both arguments with period $n^{-1}$ which reflects the fact
that as either of the spin populations moves along the channel it
experiences a periodic potential with the period set by interparticle
spacing.  Equivalently, it is a manifestation of the discreteness of the
electrons - the dot ``prefers'' to host an
integer number of electrons thus implying that the function
$U(u_{\uparrow},u_{\downarrow})$ has minima at integer values of
$nu_{\uparrow,\downarrow}$.  The classical limit of ``closed'' dot
is recovered by increasing the amplitude of $U$ to infinity while the
regime of perfect transmission is reached by decreasing the amplitude
of $U$ to zero.  For example, for the reflection amplitude
satisfying the condition $|r| \ll 1$, the form
\begin{eqnarray}
\label{perftrans}
U& =& - {\frac
{D|r|}{\pi}}(\cos2\pi nu_{\uparrow} + \cos2\pi nu_{\downarrow})\nonumber\\
& = & - {\frac {2 \gamma \Lambda_{0}\hbar|r|}{\pi}}\cos\pi \theta_{c}\cos\pi \theta_{s}
\end{eqnarray}
was derived earlier \cite{Matveev95}.  Here $D$ is the high-energy
cutoff (bandwidth) used by Matveev \cite{Matveev95} and
$\gamma = e^{\textbf{C}}$ with $\textbf{C} \approx 0.5772$ being
Euler's constant.  Shortly it will become clear why the energy and
frequency cutoffs in (\ref{perftrans}) are related by $D = \hbar
\gamma \Lambda_{0}$.

We note that without the charging term, $E_{C} = 0$,
Eqs.(\ref{effaction}) and (\ref{effpotential}) describe the dynamics
of spin-1/2
electrons in the presence of a localized inhomogeneity \cite{KF}.
Additionally, for the special case $g_{c} = g_{s}$ these equations
capture the physics of resonant tunneling of interacting \textit{spinless}
electrons onto and off a dot formed by \textit{two} tunneling barriers in the
presence of a gate which couples to the electrons between the barriers
\cite{KF}.

\section{Nearly-open dot}

Combining Eq.(\ref{effaction}), (\ref{effpotential}), and
(\ref{perftrans}) we arrive at the
effective action describing the physics of nearly-open dot
\begin{eqnarray}
\label{effactionnearlyopen}
&&A_{eff} =  \pi \hbar \sum \limits_{\nu = c,s}
\int\limits_{|\omega| \le \Lambda_{0}}  {\frac
{d\omega}{2\pi g_{\nu}}}|\omega||\theta_{\nu}(\omega)|^{2}\nonumber\\
&+&  \int d\tau  \left (E_{C}(\theta_{c} - N)^{2} - {\frac {2\gamma \Lambda_{0}
\hbar|r|}{\pi}}\cos\pi
\theta_{c}\cos\pi \theta_{s} \right )\nonumber\\
\end{eqnarray}
For almost perfect transmission $|r| \ll 1$ the action
(\ref{effactionnearlyopen}) is minimized for $\theta_{c} = N$, and the
electron number field $\theta_{c}$ can be eliminated.  This is
done by writing $\theta_{c} = N + \delta \theta_{c}$ where $\delta
\theta_{c}$ is a fluctuation and averaging
over the unperturbed, $|r| = 0$, ground state, i. e.
over the Gaussian part of (\ref{effactionnearlyopen}) involving the
fluctuating number field $\theta_{c}$:
\begin{equation}
\label{gaussianc}
A_{0c} = \int \limits _{|\omega| \le \Lambda_{0}} {\frac {d\omega}{2\pi}} ({\frac {\pi
\hbar}{g_{c}}}|\omega| + E_{c})|\theta_{c}(\omega)|^{2}
\end{equation}
This can be used to calculate the mean-square fluctuation of the number
of electrons on the dot
\begin{equation}
\label{msc}
<(\delta \theta_{c})^{2}>_{0c} = {\frac {g_{c}}{2\pi^{2}}} \ln(1 +
{\frac {\pi \hbar \Lambda_{0}}{E_{C}g_{c}}})
\end{equation}
where $<>_{0c}$ denotes an expectation value computed with $A_{0c}$,
Eq. (\ref{gaussianc}).  Similar to the spinless case \cite{Matveev95}, the
fluctuation (\ref{msc}) is finite due to the presence of the Coulomb
repulsion on the dot.  We note however that for the special case of
free electrons, $g_{c} = 2$, Eq.(\ref{msc}) agrees with
previous conclusion \cite{Matveev95} only with logarithmic accuracy.
When the frequency cutoff $\Lambda_{0}$ vanishes or the charging
energy $E_{C}$ diverges, the fluctuation (\ref{msc}) vanishes.
Additionally, when $E_{C} = 0$, the
fluctuation (\ref{msc}) diverges.  These conclusions
are in accordance with the physical expectation.

The result (\ref{msc}) allows us
to find the expectation value
\begin{equation}
\label{avcos}
<\cos \pi \theta_{c}>_{0c} = (1 + {\frac {\pi \hbar \Lambda_{0}}{E_{C}g_{c}}})^{-g_{c}/4}\cos\pi N
\end{equation}
With this in mind the action (\ref{effactionnearlyopen}) simplifies to
a reduced form involving only the spin field $\theta_{s}$
\begin{equation}
\label{effspinaction}
{\frac {A_{s}}{\hbar}} =  \pi \int\limits_{|\omega| \le \Lambda_{0}}  {\frac
{d\omega}{2\pi g_{s}}}|\omega||\theta_{s}(\omega)|^{2}
- v_{0} \Lambda_{0}\int d\tau \cos\pi \theta_{s}
\end{equation}
where the dimensionless parameter
\begin{equation}
\label{vgate}
v_{0} = {\frac {2\gamma |r|}{\pi}} (1 + {\frac {\pi \hbar \Lambda_{0}}{E_{C}g_{c}}})^{-g_{c}/4}\cos\pi N
\end{equation}
accumulates salient information about the problem at hand, and the
parameter $N$ has a meaning of the average population of the dot.

\subsection{Perturbation theory}

We start by computing a correction to the ground-state energy of the
dot $\delta E
$ perturbative in the interacting part of the action
(\ref{effspinaction})
\begin{equation}
\label{intaction}
{\frac {A_{int}}{\hbar}} = - v_{0} \Lambda_{0} \int d\tau \cos\pi \theta_{s},
\end{equation}
A quick way to arrive at corresponding expression is to notice that the action
(\ref{effspinaction}) can be formally recognized as a Hamiltonian of
a one-dimensional classical statistical mechanics problem with the
imaginary time coordinate $\tau$ corresponding to fictitious
space direction \cite{Kogut,Peierls1}.  Then the zero-point motion
is analogous to thermal fluctuations, and the result can be
inferred from the formulas of thermodynamic perturbation theory
\cite{LL1}.  Restricting ourselves to second order we have
\begin{equation}
\label{pt}
L\delta E = <A_{int}>_{0s} - {\frac{1}{2\hbar}}<(A_{int}
- <A_{int}>_{0s})^{2}>_{0s}
\end{equation}
Here $L = \hbar/T$ is the size of the fictitious classical system in the imaginary
time direction and $T$ is the temperature;  the zero-temperature
limit is equivalent to the $L \rightarrow \infty$ limit.  The notation $<>_{0s}$
indicates an
expectation value computed with the harmonic, $v_{0} = 0$, part of
$A_{s}$, Eq.(\ref{effspinaction}):
\begin{equation}
\label{harmspinaction}
{\frac {A_{0s}}{\hbar}} =  \pi \int\limits_{|\omega| \le \Lambda_{0}}  {\frac
{d\omega}{2\pi g_{s}}}|\omega||\theta_{s}(\omega)|^{2}
\end{equation}

In order to compute the first-order correction,
$\delta E_{1} = <A_{int}>_{0s}/L$, we
need to know the mean-square fluctuation of the spin field
$\theta_{s}$ which can be found to be
\begin{equation}
\label{mss}
<\theta_{s}^{2}>_{0s} = {\frac {g_{s}}{2\pi}} \int \limits _{|\omega|
\le \Lambda_{0}} {\frac {d\omega}{2\pi |\omega|}} = {\frac {g_{s}}{2\pi^{2}}} \ln(\Lambda_{0} L)
\end{equation}
This implies that
\begin{equation}
\label{firstorder}
L\delta E_{1} = <A_{int}>_{0s} = - v_{0}\hbar
(\Lambda_{0} L)^{1 - g_{s}/4},
\end{equation}
i.e. the first-order correction to the ground-state energy $\delta E_{1}
\propto L^{-g_{s}/4}$ vanishes in the zero-temperature limit.  Similar
to the free-fermion case \cite{Matveev95}, $g_{c,s} = 2$, this is a consequence of divergent spin fluctuations (\ref{mss}).

In order to calculate the second-order correction $\delta E_{2} = - <(A_{int}
- <A_{int}>_{0s})^{2}>_{0s}/2L\hbar$, we will need the correlation
function
\begin{eqnarray}
\label{corrfunction}
&\chi(\tau)& = <(\theta_{s}(\tau) -
 \theta_{s}(0))^{2}>_{0s}\nonumber\\
& = &{\frac
{g_{s}}{\pi}} \int \limits _{|\omega| \le \Lambda_{0}} {\frac
{d\omega}{2\pi}}{\frac {1 - \cos\omega \tau}{|\omega|}} = {\frac
 {g_{s}}{\pi^{2}}} {\rm Cin}(\Lambda_{0} |\tau|)
\end{eqnarray}
where ${\rm Cin}(t)$ is a cosine integral \cite{cosint}.  As $t \rightarrow 0$ the function ${\rm Cin}(t)$ vanishes as $t^{2}/4$
while as $t
\rightarrow \infty$, it diverges logarithmically, ${\rm Cin}(t)
\rightarrow \ln (\gamma t)$.  The latter asymptotics also implies the
large-argument logarithmic divergence of the correlation function (\ref{corrfunction}), $\chi(\tau) =
(g_{s}/\pi^{2})\ln(\gamma \Lambda_{0} |\tau|)$.  For noninteracting
electrons, $g_{s} = 2$, this can be compared with the result implicit
in Eq.(34) of Matveev \cite{Matveev95}:  $\chi(\tau) =
(2/\pi^{2})\ln(D|\tau|/\hbar)$.  Therefore the energy and the frequency
cutoffs, $D$ and $\Lambda_{0}$, respectively, are related by $D =
\hbar \gamma \Lambda_{0}$, which was already employed in Eq.(\ref{perftrans}).

The expression for the second-order correction $\delta E_{2}$ can be simplified to
\begin{equation}
\label{secondorder1}
\delta E_{2} = - {\frac{(v_{0}\hbar \Lambda_{0})^{2}}{2\hbar}} \int
\limits_{0}^{\infty} d\tau
\exp(-{\frac {\pi^{2} \chi(\tau)}{2}})
= - {\frac{v_{0}^{2} \hbar \Lambda_{0}}{2}} B(g_{s}),
\end{equation}
where the function
\begin{equation}
\label{bofg}
B(g_{s}) = \int \limits_{0}^{\infty} dt \exp(-{\frac {g_{s}}{2}} {\rm Cin}(t))
\end{equation}
is only defined for $g_{s} > 2$.  The
evaluation of the integral in Eq.(\ref{bofg}) employing the
logarithmic approximation of the cosine integral,  ${\rm Cin}(t \le \gamma^{-1}) = 0$ and ${\rm Cin}(t >
\gamma^{-1}) = \ln(\gamma t)$, gives
\begin{equation}
\label{bofgappr}
B(g_{s}) \approx {\frac{g_{s}}{\gamma(g_{s} - 2)}}
\end{equation}
Although this estimate becomes asymptotically exact as $g_{s}
\rightarrow 2$ from above, as $g_{s}$ increases, its accuracy
degrades.  Specifically, as $g_{s} \rightarrow \infty$, the integral
(\ref{bofg}) is dominated by the region of small $t$ where the
logarithmic approximation is not adequate and we find instead $B(g_{x}) \approx
(2\pi/g_{s})^{1/2}$.  However exceedingly large values of the spin
correlation exponent $g_{s}$ are non-physical and will not be
considered.

In order to evaluate a low-temperature contribution into $\delta
E_{2}$ we need to replace
the upper integration limit in
Eq.(\ref{bofg}) with $\Lambda_{0} L/2 \gg 1$.  As a result we find
\begin{eqnarray}
\label{secondorder2}
\delta E_{2}& =& - {\frac{v_{0}^{2} \hbar \Lambda_{0}}{2}} B(g_{s})\left
(1 -
{\frac {2(\gamma \Lambda_{0}L/2)^{1 - g_{s}/2}}{\gamma (g_{s} -
2)B(g_{s})}}\right )\nonumber\\
& \approx &  - {\frac{v_{0}^{2} \hbar \Lambda_{0}}{2}} B(g_{s})\left
(1 -
{\frac {2(\gamma \Lambda_{0}L/2)^{1 - g_{s}/2}}{g_{s}}}\right )
\end{eqnarray}
where in the last step we used the approximation (\ref{bofgappr}).  For $g_{s} > 2$ we may take the limit of zero temperature, $L
\rightarrow \infty$, which takes us back to Eq.(\ref{secondorder1}),
while in the presence of $SU(2)$ spin symmetry, $g_{s} = 2$, we find
with logarithmic accuracy
\begin{equation}
\label{secondorderg_s=2}
\delta E_{2} =  - {\frac{v_{0}^{2} \hbar \Lambda_{0}}{2\gamma}} \ln(\Lambda_{0} L)
\end{equation}
In the zero-temperature limit, $L \rightarrow \infty$, this
result diverges which means that a non-perturbative
calculation is required.  The logarithmic divergence
(\ref{secondorderg_s=2}) is analogous to its free-fermion counterpart
$g_{c,s} = 2$ found earlier \cite{Matveev95}.

Additionally, the
special place of the limit of non-interacting electrons can be seen by taking the
continuum limit, $\Lambda_{0} \rightarrow \infty$, in Eq. (\ref{secondorderg_s=2}).  This only leaves the
logarithmic dependence on the microscopic cutoff,
$\delta E_{2} = - (4\gamma
|r|^{2}E_{C}/\pi^{3})\ln(\Lambda_{0}L)$, thus implying that the
prefactor is solely determined by macroscopic quantities.

\subsection{Beyond perturbation theory: heuristic argument}

In order to go beyond the perturbation theory, first we need to clarify its
range of applicability.  A naive expectation that the
perturbative result (\ref{secondorder1})  is reliable for $g_{s} >
2$ as long as $|v_{0}| \ll 1$ is
misleading.  To find the correct condition we adopt an argument due to
J. Lajzerowicz originally given in the classical context of the
roughening phase transition \cite{Nozieres}:

The cumulant expansion (\ref{pt}) is performed in powers of
$A_{int}/\hbar$ which implies that the perturbation theory is
applicable as long as $|<A_{int}>_{0s}/\hbar |\ll 1$.  Inspection of
Eq.(\ref{firstorder}) shows that this will be the case only for $g_{s}
\ge 4$ and $|v_{0}| \ll 1$.  One can also arrive at the same conclusion by applying the standard
argument \cite{LL1} which compares the magnitudes of the first and
second-order terms of the expansion (\ref{pt}).

The physics in the non-perturbative regime $2 \le g_{s} < 4$ can be
understood heuristically by noticing that the perturbative treatment
assumes that spin fluctuations grow with scale without bound (see
Eq.(\ref{mss})) which is indeed the case for $g_{s} \ge 4$.  This
assumption, however, breaks down for $g_{s} < 4$ because
spin fluctuations are suppressed at scales exceeding a characteristic value
$L_{c}$ which has a meaning of the correlation length of the
fictitious classical statistical mechanics problem corresponding to the action
(\ref{effspinaction}).  This scale marks the verge of
applicability of the perturbation theory, $|<A_{int}/\hbar| \simeq 1$,
which with the help of Eq.(\ref{firstorder}) leads to the estimate
\begin{equation}
\label{corlength}
L_{c} \simeq \Lambda_{0}^{-1} |v_{0}|^{-4/(4 - g_{s})}
\end{equation}
In the limit $g_{s} \rightarrow 4$ from below the correlation length
(\ref{corlength}) diverges
which is in correspondence with the well-known fact \cite{SG} that the
model (\ref{effspinaction}) has a phase transition
at $g_{s} = 4$.  We also note that for the case of
non-interacting electrons, $g_{c,s} = 2$, the energy scale
$\hbar/L_{c}$ corresponding to (\ref{corlength}) has been identified
earlier \cite{Aleiner}; a related renormalization-type argument
is given in Section IIIC.

At the scales exceeding $L_{c}$ the spin
fluctuations are effectively frozen and no longer contribute into the
ground-state energy.  Thus
the latter can be estimated by substituting $L \simeq L_{c}$ in
Eqs.(\ref{secondorder2}) and (\ref{secondorderg_s=2}).  As a
result for $2 < g_{s} < 4$ the correction to the ground-state energy
$\delta E$ is estimated as
\begin{equation}
\label{estimateg_s>2}
\delta E \simeq  - {\frac{v_{0}^{2} \hbar \Lambda_{0}}{2}} B(g_{s})\left
(1 -
{\frac {|v_{0}|^{2(g_{s} - 2)/(4 - g_{s})}}{g_{s}}} \right )
\end{equation}
while for $g_{s} = 2$ we find
\begin{equation}
\label{estimateg_s=2}
\delta E =  - {\frac{v_{0}^{2} \hbar \Lambda_{0}}{2\gamma}}
\ln\left ({\frac {1}{v_{0}^{2}}}\right )
\end{equation}
The numerical uncertainty present in the definition of the characteristic
scale $L_{c}$, Eq. (\ref{corlength}), translates into an
uncertainty of the term of Eq. (\ref{estimateg_s>2})
non-analytic in the parameter $v_{0}$:  here  $v_{0}$ should be viewed as
$v_{0}$ times a constant of order unity.  However the analytic $v_{0}^{2}$
term is given exactly.  At the same time
Eq.(\ref{estimateg_s=2}) continues to possess logarithmic accuracy as
the uncertainty of $L_{c}$ only affects the argument of the logarithm.

Since the parameter $v_{0}$, Eq.(\ref{vgate}), vanishes at half-integer
dot population $N$, the ground-state energy given by
Eqs. (\ref{estimateg_s>2}) and (\ref{estimateg_s=2}) and
related equilibrium properties exhibit non-analytic behavior at
half-integer $N$.  For example for non-interacting electrons, $g_{c,s}
= 2$ (or more generally in the presence of $SU(2)$ spin symmetry,
$g_{s} = 2$), the capacitance of the dot $C \propto \partial^{2}
E/\partial N^{2}$ diverges logarithmically at half-integer dot
population, the result derived by Matveev through an exact calculation
\cite{Matveev95}.

More generally, $2 < g_{s} < 4$, the
singular part of the energy behaves as $\delta E_{sing} \propto
|\delta N|^{4/(4 - g_{s})}$ (see Eq.(\ref{estimateg_s>2})) where $\delta N$ is a small deviation from
a half-integer dot population.  Although the corresponding contribution into
the capacitance $C_{sing} \propto |\delta N|^{2(g_{s} - 2)/(4 -
g_{s})}$ vanishes, the third-order derivative of the energy, $\partial^{3}
E_{sing}/\partial N^{3} \propto |\delta N|^{(3g_{s} - 8)/(4 - g_{s})}$
diverges for $g_{s} < 8/3$.  Generally, the spin correlation
exponent $g_{s}$ parameterizes the order of the
phase transition at half-integer population of the dot
(i.e. the order of the first
divergent derivative of the ground-state energy):  it is of the second-order at
$g_{s} =2$ and of the $l$th-order for $g_{s} < 4(l -
1)/l$ and  $l > 2$.  As $g_{s}$ increases from $2$ to $4$,
spin fluctuations increase, the non-analyticity of the ground-state
energy manifests itself in the divergence of progressively
higher-order derivative and the order of the phase transition
increases approaching infinity at $g_{s} = 4$.  The ultimate 
origin of these singularities is the discreteness of the electron
spin.  For $g_{s} \ge 4$ spin fluctuations become unbound and the
correction to the ground-state energy becomes regular and proportional to
$-(\cos2\pi N + 1)$ (see Eqs.(\ref{secondorder1}) and
(\ref{vgate})) still exhibiting remnants of the discreteness of the
electrons.

\subsection{Beyond perturbation theory:  renormalization-group analysis}

The heuristic results (\ref{estimateg_s>2}) and (\ref{estimateg_s=2})
can be placed on a firmer ground through a perturbative, $v_{0} \ll 1$,
renormalization-group analysis of the effective action 
(\ref{effspinaction}).  It is well-known that renormalization
procedure involves repeated elimination of microscopic degrees of
freedom \cite{Ma}.  This process generates
terms which renormalize the parameters of the initial action thus
bringing about observable physics, and
contributions which are irrelevant in the renormalization-group sense
and thus can be dropped.  For the model (\ref{effspinaction})
the associated terms are discussed in the literature \cite{SG}.
Additionally, elimination of microscopic degrees of freedom can generate
extensive quantities which (in the present context) represent contributions into the
ground-state energy.  As far as we know, these contributions which
will be our primary concern, have not been discussed which 
gives us an excuse to summarize the salient steps of the whole calculation.   

In the sequel we adopt the approach of Nozi{\`e}res \cite{Nozieres}
who investigated the
classical problem of the roughening phase transition.  As usual, the field $\theta_{s}$ is presented as a sum of ``fast'',
$\theta_{s}^{(f)}$, and ``slow'', $\theta_{s}^{(s)}$, modes:
\begin{equation}
\label{fastslow}
\theta_{s}(\tau) = \theta_{s}^{(f)}(\tau) + \theta_{s}^{(s)}(\tau)
\end{equation}
Below we will use the sharp cutoff procedure
\cite{Ma} appropriate for the model (\ref{effspinaction}), i. e. choose
\begin{equation}
\label{fastcutoff}
\theta_{s}^{(f)}(\omega) = \theta_{s}(\omega), ~~~~\Lambda_{0}(1 - \epsilon) < |\omega| \le \Lambda_{0}, 
\end{equation}
where $\epsilon \ll 1$, and
\begin{equation}
\label{slowcutoff}
\theta_{s}^{(s)}(\omega) = \theta_{s}(\omega). ~~~~|\omega| \le
\Lambda_{0}(1 - \epsilon)
\end{equation}
After the fast degrees of freedom are eliminated, the
effective action (\ref{effspinaction}) turns into a functional of slow
modes, $A_{s}[\theta_{s}^{(s)}(\tau)]$, which to second order in
$v_{0}$ is given by the expansion \cite{Nozieres,SG}:
\begin{eqnarray}
\label{cumexpansion}
&&A_{s}[\theta_{s}^{(s)}(\tau)] = \pi \hbar \int\limits_{|\omega| \le
\Lambda_{0}(1 - \epsilon)}  {\frac
{d\omega}{2\pi g_{s}}}|\omega||\theta_{s}(\omega)|^{2}\nonumber\\& 
+ 
& <A_{int}[\theta_{s}^{(s)}(\tau) +
\theta_{s}^{(f)}(\tau)]>_{f}\nonumber\\& 
- &{\frac{1}{2\hbar}}<(A_{int}[\theta_{s}^{(s)}(\tau) +
\theta_{s}^{(f)}(\tau)]\nonumber\\& 
-&<A_{int}[\theta_{s}^{(s)}(\tau) +
\theta_{s}^{(f)}(\tau)]>_{f} )^{2}>_{f}
\end{eqnarray} 
Here $<>_{f}$ denotes averaging over fast degrees of freedom with 
the Gaussian probability distribution determined by the action (\ref{harmspinaction}) except that the
frequency range, according to Eq.(\ref{fastcutoff}), is $\Lambda_{0}(1 - \epsilon) < |\omega| \le \Lambda_{0}$.

To first order we find
\begin{eqnarray}
\label{firstorderrg1}
A_{1}& = & - v_{0}(1 - {\frac
{\pi^{2}}{2}}<(\theta_{s}^{(f)})^{2}>_{f})\Lambda_{0} \hbar \int d\tau
\cos\pi \theta_{s}^{(s)}\nonumber\\
&=&  - v_{0}(1 - {\frac
{g_{s}}{4}}\epsilon)\Lambda_{0} \hbar \int d\tau
\cos\pi \theta_{s}^{(s)}\nonumber\\
&=&  - v\Lambda \hbar \int d\tau
\cos\pi \theta_{s}^{(s)} ,
\end{eqnarray} 
where we employed the fact that the mean-square fluctuation of the fast modes,
$<(\theta_{s}^{(f)})^{2}>_{f}$, is given by the integral (\ref{mss}) 
except that the integration range is $\Lambda_{0}(1 - \epsilon) < |\omega| \le
\Lambda_{0}$, Eq.(\ref{fastcutoff}).  The last representation in
(\ref{firstorderrg1}) combined with the first term of
(\ref{cumexpansion}) form an action having the same functional 
form as the original action $A_{s}$, Eq.(\ref{effspinaction}) except
for smaller cutoff, 
\begin{equation}
\label{newlambda}
\Lambda = \Lambda_{0}(1 - \epsilon),
\end{equation} 
and a slightly different amplitude of the cosine term,
\begin{equation}
\label{newv}
v = v_{0} {\frac {1 - g_{s}\epsilon/4}{1 - \epsilon}},
\end{equation} 

To second order we obtain
\begin{eqnarray}
\label{secondorderrg1}
A_{2} =& -& {\frac {v_{0}^{2}\Lambda_{0}^{2} \pi^{2}\hbar}{2}} \int d\tau
d\tau' \sin\pi \theta_{s}^{(s)}\sin\pi \theta_{s}^{(s)}~'\nonumber\\
&\times&<\theta_{s}^{(f)}\theta_{s}^{(f)}~'>_{f},
\end{eqnarray}
where the prime means that the argument of corresponding function is
$\tau'$.  The correlation function
$<\theta_{s}^{(f)}\theta_{s}^{(f)}~'>_{f}$ is given by
\begin{eqnarray}
\label{fastcorrfunction}
<\theta_{s}^{(f)}\theta_{s}^{(f)}~'>_{f}& =& {\frac {g_{s}}{2\pi}} \int
\limits_{\Lambda_{0}(1 - dl) < |\omega| \le \Lambda_{0}} {\frac
{d\omega}{2\pi}} {\frac {\exp i \omega(\tau -
\tau')}{|\omega|}}\nonumber\\
&=& {\frac {g_{s}}{2\pi^{2}}} \epsilon \cos(\Lambda_{0}(\tau - \tau')),
\end{eqnarray}  
which allows us to rewrite Eq.(\ref{secondorderrg1}) in a more
explicit form:
\begin{eqnarray}
\label{secondorderrg2}
A_{2}& = &{\frac {v_{0}^{2}\Lambda_{0}^{2}\hbar g_{s}}{8}} \epsilon \int d\tau
d\tau' \cos(\Lambda_{0}(\tau - \tau'))\nonumber\\
&\times& \left (\cos\pi(\theta_{s}^{(s)} + \theta_{s}^{(s)}~') -
\cos\pi(\theta_{s}^{(s)} - \theta_{s}^{(s)}~')\right )  
\end{eqnarray}

A further truncation is necessary to see the implications of
Eq.(\ref{secondorderrg2}).  First, the $\cos\pi(\theta_{s}^{(s)} +
\theta_{s}^{(s)}~')$ term can be dropped since for $\tau = \tau'$ it
becomes the second harmonic of the main $\cos\pi\theta_{s}^{(s)}$
oscillation, and thus is irrelevant in the renormalization-group sense.   

Gradient expansion of the other term 
\begin{eqnarray}
\label{gradientwrong}
\cos\pi(\theta_{s}^{(s)} - \theta_{s}^{(s)}~') &\approx& 1 - {\frac
{\pi^{2}}{2}} (\theta_{s}^{(s)} - \theta_{s}^{(s)}~')^{2}\nonumber\\
& \approx& 1 - {\frac
{\pi^{2}}{2}} ((\tau - \tau')d\theta_{s}^{(s)}/d\tau)^{2} 
\end{eqnarray} 
is not appropriate for large magnitude of the difference $\tau -
\tau'$ where the two values $\theta_{s}^{(s)}$ and
$\theta_{s}^{(s)}~'$ are hardly correlated:  the right-hand side
of (\ref{gradientwrong}) needs a factor specifying the extent to
which these values are correlated.   The same situation is encountered
in the context of the usual sine-Gordon theory where Knops and den
Ouden \cite{Knops} demonstrated that continuing the expansion of the
cosine to all orders generates the necessary factor.  Below we adopt
an equivalent physical argument due to Nozi{\`e}res \cite{Nozieres},
and go beyond the existing treatments of the action (\ref{effspinaction}):

The field $\theta_{s}^{(s)}$ in (\ref{gradientwrong}) should be considered
as a sum of the equilibrium fluctuation $\theta_{s}^{(s)}$ and a small
correction $\zeta$ (which might be a response to a small low-frequency
perturbation).  Thus instead of (\ref{gradientwrong}), first we average
over equilibrium fluctuations which is followed by the expansion in
$\zeta$.  This generates the expansion replacing
Eq.(\ref{gradientwrong}):
\begin{eqnarray}
\label{gradientright}
&&\cos\pi(\theta_{s}^{(s)} - \theta_{s}^{(s)}~')\nonumber\\
 &\rightarrow&
(1 - {\frac
{\pi^{2}}{2}} (\zeta - \zeta')^{2})<\cos\pi(\theta_{s}^{(s)} - \theta_{s}^{(s)}~')>_{0s}\nonumber\\
&\approx& \left (1 - {\frac
{\pi^{2}}{2}} ((\tau - \tau')d\theta_{s}^{(s)}/d\tau)^{2}\right )\nonumber\\
&\times& \exp(-{\frac {\pi^{2} \chi(\tau - \tau')}{2}})
\end{eqnarray}
We note that the same correlation exponential appears in the second-order
result, Eq.(\ref{secondorder1}).
>From technical viewpoint, the
presence of this factor eliminates all the potential difficulties
usually encountered when sharp cutoff procedures are used \cite{Ma,Kogut}.

Insertion of the expansion (\ref{gradientright}) back into (\ref{secondorderrg2})
makes it clear that a small positive contribution of
the $\int d\tau (d\theta_{s}^{(s)}/d\tau)^{2}$ form is generated under
renormalization.  For the classical problem of the roughening phase
transition such contributions are important parts of the physics as
they renormalize the interface stiffness \cite{Nozieres}.  However in the
present context they are irrelevant and thus can be
dropped \cite{SG}.  For the model (\ref{effspinaction}) the role of
the stiffness is played by $g_{s}^{-1}$ which does not renormalize to
any order \cite{SG}:  renormalization procedure cannot generate non-analytic
contributions proportional to $|\omega|$.

The constant term in (\ref{gradientright}) generates a small negative
correction to the ground-state energy
\begin{equation}
\label{secondorderrg}
\Delta E = - {\frac {v_{0}^{2} \Lambda_{0} \hbar}{2}}H(g_{s})\epsilon,
\end{equation}
where the function
\begin{equation}
\label{hofg1}
H(g_{s}) = {\frac {g_{s}}{2}}\int \limits_{0}^{\infty} dt \cos t \exp (-{\frac {g_{s}}{2}} {\rm Cin}~(t))
\end{equation}
is defined unambiguously for $g_{s} \ge 2$.  Using the definition of
the cosine integral (\ref{corrfunction}) this formula can be
rewritten as
\begin{eqnarray}
\label{hofgs}
H(g_{s})& = &{\frac {g_{s}}{2}}\int \limits_{0}^{\infty} \left (1 - t {\frac {d({\rm
Cin}(t))}{dt}}\right )\exp (-{\frac {g_{s}}{2}} {\rm Cin}(t))dt\nonumber\\
&=& \int \limits_{0}^{\infty} t^{1 - g_{s}/2} d\left (t^{g_{s}/2}\exp
(-{\frac {g_{s}}{2}} {\rm Cin}(t))\right )
\end{eqnarray}
which implies
\begin{equation}
\label{dof2}
H(2) = \lim_{t \rightarrow \infty} (t \exp
(-{\rm Cin}(t))) = \gamma^{-1} \approx 0.5615
\end{equation}
This demonstrates the special role played by $g_{s} = 2$ - only
in the presence of $SU(2)$ spin symmetry is the corresponding value
of $H(g_{s})$ controlled by the $t \rightarrow \infty$ limit.

For $g_{s} > 2$ we integrate Eq.(\ref{hofgs}) by parts which
establishes a relationship with the function $B(g_{s})$,
Eq.(\ref{bofg}), appearing in the second-order expression
(\ref{secondorder1}):
\begin{equation}
\label{hvsb}
H(g_{s}) = ({\frac {g_{s}}{2}} - 1) B(g_{s})
\end{equation}

Renormalization-group transformation which led to
Eqs.(\ref{newlambda}), (\ref{newv}), and (\ref{secondorderrg}) can be
carried over repeatedly by reducing the frequency cutoff by successive
infinitesimally small amounts $\epsilon = dl \rightarrow 0$.  Since each step of the transformation
incorporates the effect of previous renormalizations, the outcome is
customary to represent in the form of differential flow equations:
\begin{equation}
\label{flambda}
{\frac {d\Lambda}{dl}} = - \Lambda
\end{equation}
\begin{equation}
\label{fv}
{\frac {dv}{dl}} = (1 - {\frac {g_{s}}{4}})v
\end{equation}
\begin{equation}
\label{fenergy1}
{\frac {dE}{dl}} = - {\frac {v^{2} \Lambda \hbar}{2}}H(g_{s})
\end{equation}
These equations have to be solved subject to the ``initial''
conditions, $\Lambda(l = 0) = \Lambda_{0}$, $v(l = 0) = v_{0}$;
the macroscopic behavior is recovered in the $\Lambda(l \rightarrow
\infty) \rightarrow 0$ limit.

Eq.(\ref{flambda}) whose solution is given by
\begin{equation}
\label{lambdaofl}
\Lambda(l) = \Lambda_{0} e^{-l}
\end{equation}
is a common place in renormalization-group calculations.

Eq.(\ref{fv}) and its solution
\begin{equation}
\label{vofl}
v(l) = v_{0} e^{(1 - g_{s}/4)l}
\end{equation}
are also well-known \cite{SG} from renormalization-group treatments of the model
(\ref{effspinaction}):

For $g_{s} > 4$ the parameter $v(l)$
flows to zero as $l \rightarrow \infty$ thus implying that the last term of
(\ref{effspinaction}) is irrelevant in renormalization-group sense.
This is the region of applicability of perturbative approach with
unbound spin fluctuations on the dot (see Eq.(\ref{mss})).  For $g_{s}
< 4$ the parameter $v(l)$ grows under renormalization indicating that
the discreteness of the electron spin is relevant and that
perturbative treatment fails at low frequencies.  As a result the
fluctuations of the spin of the dot will be suppressed at flow
parameters exceeding a characteristic value $l_{c}$ such as $|v(l_{c})|
\simeq 1$:
\begin{equation}
\label{lc}
l_{c} \simeq {\frac {4} {4 - g_{s}}} \ln\left ({\frac
{1}{|v_{0}|}}\right )
\end{equation}
We note that the corresponding scale in imaginary time
direction $\Lambda_{c}^{-1} = \Lambda_{0}^{-1} e^{l_{c}}$ coincides
with the heuristic result (\ref{corlength}) - the argument that led to
(\ref{corlength}) is equivalent to the first-order
renormalization-group calculation.

Eq.(\ref{fenergy1}) describing
renormalization of the ground-state energy due to eliminated
high-frequency modes is novel; its counterpart
in the context of the conventional sine-Gordon problem is known
\cite{Kogut,Nozieres}.   In order to compute the contribution into the ground-state energy we
first substitute the solutions (\ref{lambdaofl}) and (\ref{vofl}) in
Eq.(\ref{fenergy1}) which turns it into a form
\begin{equation}
\label{fenergy2}
{\frac {dE}{dl}} = - {\frac {v_{0}^{2} \Lambda_{0} \hbar}{2}}H(g_{s})
e^{(1 - g_{s}/2)l}
\end{equation}
that makes explicit a special role played by $SU(2)$ spin symmetry, $g_{s} =
2$.

For $g_{s} \ge 4$ spin fluctuations of all frequencies contribute into
the ground-state energy thus implying that Eq.(\ref{fenergy2}) should
be integrated between zero and infinity.  Combined with
Eq.(\ref{hvsb}), it then reproduces the perturbative result
(\ref{secondorder1}).

For $2 \le g_{s} < 4$ spin fluctuations are suppressed at small
frequencies and the contribution into the ground-state energy can be
found by integrating Eq.(\ref{fenergy2}) between zero and the characteristic
flow parameter $l_{c}$, Eq.(\ref{lc}).  As a result for $2 < g_{s} <
4$ we find
\begin{equation}
\label{rgenergy2<g<4}
\delta E \simeq - {\frac {v_{0}^{2} \Lambda_{0} \hbar}{2}} B(g_{s}) (1
- |v_{0}|^{2(g_{s} - 2)/(4 - g_{s})})
\end{equation}
This is very similar to the heuristic result (\ref{estimateg_s>2}),
and the uncertainty of the second term of (\ref{rgenergy2<g<4}) having
its origin in the uncertainty built into the definition of the
characteristic flow parameter $l_{c}$, Eq.(\ref{lc}), makes
Eqs.(\ref{estimateg_s>2}) and (\ref{rgenergy2<g<4}) practically
indistinguishable.

Finally, for $g_{s} = 2$ integrating Eq.(\ref{fenergy2}) between zero
and $l_{c}$, Eq.(\ref{lc}), and taking into account the numerical
value $H(2)$, Eq.(\ref{dof2}), reproduces the heuristic result
(\ref{estimateg_s=2}).

\subsection{Analogy to the sine-Gordon theory}

The main conclusion of previous analysis is that in the free-fermionic/$SU(2)$
symmetric limit, $g_{s}
= 2$, the ground-state energy (as a function of deviation $\delta N$
away from half-integer dot population) exhibits logarithmic non-analyticity,
$\delta E_{sing} \propto  \delta N^{2} \ln(1/|\delta N|)$;  in the
interacting case, $2 < g_{s} < 4$, this is replaced by a power law,
$\delta E_{sing} \propto  |\delta N|^{4/(4 - g_{s})}$.  This outcome is
not unique to the zero-dimensional field theory (\ref{effspinaction}) describing the dynamics of
nearly-open quantum dot - the conventional one-dimensional sin-Gordon
field theory shows the same effect.  The pertinent physical
phenomenon here is the Peierls instability \cite{Peierls}:

More than fifty years ago Peierls demonstrated that a one-dimensional
metal is subject to a spontaneous lattice distortion that converts it
into an insulator.  The distortion causes commensuration of electronic
and ionic subsystems which leads to the reduction of the electronic part of
the system energy.  For non-interacting electrons Peierls found that
the energy gain of commensuration as a function of small lattice distortion
$\delta$ has non-analytic form $\delta E_{sing} \propto \delta^{2} \ln(1/
\delta)$.   Since other effects favoring regular lattice spacing
$\delta = 0$ increase the energy by an amount proportional to
$\delta^{2}$, the distortion is inevitable.

Since the Peierls phenomenon just barely occurs for non-interacting
electrons, interelectron interactions have a drastic
effect on the fate of the instability.  For spinless electrons the
phenomenological theory of the effect which takes into account
interactions, has been given in the past \cite{KS1}.  There it was
demonstrated that the electronic energy gain due to the lattice
distortion is given by the ground-state energy corresponding to the sine-Gordon action
\begin{equation}
\label{sgaction}
A = \int dxd\tau \left ({\frac {\rho}{2}} (({\frac {\partial u}{\partial \tau}})^{2} +
c^{2}({\frac {\partial u}{\partial x}})^{2}) - \lambda \cos2\pi nu
\right )
\end{equation}
Here $u(x, \tau)$ is the electron displacement field, $\rho$ is the
mass density, $c$ is the sound velocity, $n$ is the number density,
and $\lambda$ is proportional to the amplitude of the lattice
distortion $\delta$.   The
effect of interelectron interactions can be parameterized by
dimensionless combination
\begin{equation}
\label{spinlessg}
g = {\frac {\pi \hbar n^{2}}{\rho c}}
\end{equation}
which can be recognized as the ``spinless'' counterpart of the charge
correlation exponent $g_{c}$ (\ref{exponents}).  Renormalization-group
analysis of the action (\ref{sgaction}) resembling that of
previous subsection then leads to the following conclusions \cite{KS1}:

For non-interacting electrons, $g = 1$, the form,  originally derived by Peierls  \cite{Peierls},
$\delta E_{sing} \propto  \delta^{2} \ln(1/\delta)$, is found.
This is
analogous to the ``non-interacting'' result,
Eq.(\ref{estimateg_s=2}).

In the presence of interactions satisfying $g < 2$
(and excluding $g = 1$), the singular part of the ground-state
energy as a function of the lattice distortion $\delta$ is given by
$\delta E_{sing} \propto  \delta^{2/(2 - g)}$.  This is analogous to
our result $\delta E_{sing} \propto  |\delta N|^{4/(4 - g_{s})}$
(see Eqs. (\ref{estimateg_s>2}) and (\ref{rgenergy2<g<4})).

One of the important differences between the actions
(\ref{effspinaction}) and (\ref{sgaction}) is that the spin
correlation exponent $g_{s}$ (\ref{exponents}) controlling the degree
of zero-point motion in (\ref{effspinaction}) does not renormalize.
For the sin-Gordon action (\ref{sgaction}) the same role is played by
the exponent $g$ (\ref{spinlessg}) which does renormalize.  However
for $g < 2$ and small distortion (small $\lambda$) this renormalization is negligible
\cite{KS1}.  That is why an analogy can be drawn between the
ground-state properties of the field theories (\ref{effspinaction}) and
(\ref{sgaction}).

\section{Nearly-closed dot}

Although the analysis of
Section III makes it clear that the logarithmic divergence of the
capacitance at half-integer dot population
\cite{Matveev95} is a consequence of assumption that the electrons in
the constriction are non-interacting, it cannot explain why the same
type of singularity may be present in the weak-tunneling regime
\cite{Matveev91} and possibly for any tunneling \cite{Matveev95}.  To
address these issues, we turn our attention to the limit of
a nearly-closed quantum dot.

The starting point of the analysis is clarification of the nature of the classical
ground state described by the effective action (\ref{effaction}).  This is
determined by the minima of the potential $V(\theta_{c}, \theta_{s})$,
Eq.(\ref{effpotential}), and has been discussed by Kane and Fisher
\cite{KF} in a related context of resonant tunneling of spinless
electrons confined between two barriers in the presence of a gate.
Although the potential entering Eq.(\ref{effactionnearlyopen}) is not
quantitatively correct for a nearly-closed dot, it nevertheless retains
the symmetries of the problem and can be used as an illustration
of the argument below disregarding the $|r| \ll 1$ constraint.  Visual
aid is provided by Fig.
1 which parallels Figure 2 of Kane and Fisher \cite{KF}.

Without the charging term, $E_{c} = 0$, the classical ground-state
is degenerate with infinitely sharp minima located at integer values of the dot
population $\theta_{c}$, which reflects the discreteness of the
electrons.  The charging term of (\ref{effpotential}) lifts this
degeneracy so that for most values of the dimensionless gate voltage
$N$ there is a unique ground state determined by an integer
$\theta_{c}$ closest to $N$ with energy gaps of the order $E_{C}$ or
larger to the states of higher or lower dot population.

\begin{figure}
\includegraphics[
  width=1.0\columnwidth,
  keepaspectratio]{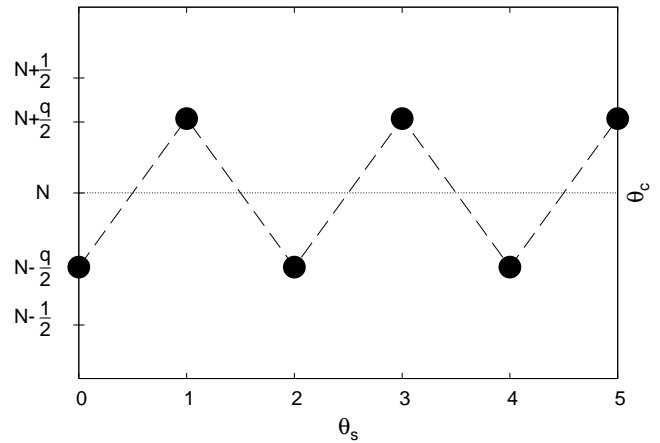}
\caption{Positions of the degenerate minima of the potential
$V(\theta_{c},\theta_{s})$ entering
Eq.(\ref{effaction}) in the $(\theta_{c},\theta_{s})$ plane for
half-integer dimensionless gate voltage $N$.  For the
purpose of illustration, nearest minima are connected by dashed line
segments.  The parameter $q$ has a
meaning of the distance between the minima in the $\theta_{c}$
direction.  Although $q = 1$ for a nearly-closed dot, zero-point motion
renormalizes it to a $0 \le q < 1$ value.  In the drawing we assumed
(without the loss of generality) that $N + 1/2$ is odd.}
\end{figure}

For half-integer values of the dimensionless gate voltage $N$ the ground state is doubly degenerate.  Since
all other integer population states are separated from these two by
energy gaps, it seems plausible \cite{Matveev91,KF} that they can be
eliminated from the discussion of the hopping dynamics when the dot is
slightly open and charge and spin fluctuations are included.  The
validity of this two-state approximation will be discussed below but
for now we assume that it is justified.  Then the $\theta_{c}$
coordinates of the minima of the effective potential $V(\theta_{c},
\theta_{s})$ are either $N + 1/2$ or $N - 1/2$ while their
$\theta_{s}$ positions are integers with the constraint that for every
minimum its $\theta_{c}$ and $\theta_{s}$ coordinates have the
same parity.  This means that if nearest minima are connected by
straight line segments (see Fig. 1), they will form an
infinite zigzag running on average parallel to the $\theta_{s}$ axis.

Opening the dot means that the barriers separating degenerate
states of integer population are no longer infinitely high - the effective
potential will now have degenerate minima whose $\theta_{c}$ positions
are either $N + q/2$ or $N - q/2$ where $q$ is the distance
between the minima of $V$ along the $\theta_{c}$ axis \cite{note2}.  At the same
time the step of the zigzag in the $\theta_{s}$ direction which is the
symmetry of the problem remains unchanged.  Including
charge and spin fluctuations, i. e. allowing finite $g_{c,s}$,
Eq. (\ref{exponents}), has the same effect of downward renormalization
of the distance between the minima of $V$ along
the $\theta_{c}$ direction.  Indeed, even for a nearly-closed quantum dot where
$q = 1$ initially, quantum-mechanical tunneling
between nearest degenerate ground-states reduces it to a $0 \le q < 1$
value.  For $q = 0$ the effective potential energy evaluated at
$\theta_{c} = N$ is $\theta_{s}$-independent.  We note that observable value of the parameter $q$ corresponds
to the jump of the dot population as the dimensionless gate voltage $N$ is swept
across a half-integer value. Then $q = 0$ means that the Coulomb
staircase is smeared.

The elementary act of tunneling between nearest ground states consists of a change of the
$\theta_{c}$ coordinate by $q$ accompanied by simultaneous change of
$\theta_{s}$ by unity.  The zigzag sequence of the ground states
implies that the tunneling event $\Delta \theta_{c} = \pm q$ must be
followed by another tunneling event consisting in opposite sign change
of $\Delta \theta_{c} = \mp q$.  At the same time tunneling
changes of $\theta_{s}$ by unity are not correlated.

To gain better understanding of the physics in the regime of a nearly-closed
quantum dot we invoke a mapping \cite{Kogut} between zero-dimensional quantum field
theory defined by the action (\ref{effaction}) and that of one-dimensional
classical statistical mechanics.  Similar to the observation made for
the spinless version of the problem \cite{KKQ},
Eq. (\ref{effaction}) can be viewed as an effective Hamiltonian of
one-dimensional
classical statistical mechanics problem involving fluctuating
two-component field $\theta_{c,s}$ with imaginary time $\tau$ playing a
role of fictitious space direction.  The $|\omega|$ dependence of the
first term of Eq. (\ref{effaction}) implies a $1/\tau^{2}$
interaction between the fields $\theta_{c,s}$ separated by distance
$\tau$.

If not for the presence of the spin field $\theta_{s}$, then for
half-integer $N$ the classical
statistical mechanics problem in question would be identical to that
of a ferromagnetic Ising model with $1/\tau^{2}$ interactions in
zero magnetic field.  Similar to the Ising model there could be two
phases:

In the ordered (broken symmetry) phase the system, initially prepared
in one of the minima of the potential $V(\theta_{c}, \theta_{s})$, will
reside in there indefinitely long despite occasional excursions
induced by thermal fluctuations into neighboring minima followed by
returns into the original minimum.  The excursion is a nucleus of
neighboring ground state constricted by two domain walls having
opposite signs of the change $\Delta \theta_{c}$.   For the Ising model with
$1/\tau^{2}$ interactions the domain walls are known to be responsible for the destruction of order
\cite{Thouless};  the same is expected to hold for the model
(\ref{effaction}).  Domain walls have their quantum-mechanical
counterparts in instantons - extremal solutions of the equations of
classical dynamics in imaginary time satisfying the condition that as
imaginary time $\tau$ traverses from minus infinity to infinity, the instanton
interpolates between degenerate ground states.

The similarity to the Ising model also implies that the broken
symmetry phase of (\ref{effaction}) may be removed by proliferating
domain walls.  As a result the system will freely hop through the minima of
$V(\theta_{c}, \theta_{s})$ leading to the
destruction of the Coulomb staircase in the quantum-mechanical context.

In order to understand quantitatively the role of the domain walls we
find it useful to rewrite the action (\ref{effaction}) in an
equivalent form which abandons the Fourier representation of the first
term and shows explicitly that the latter vanishes for the uniform system:
\begin{eqnarray}
\label{cgeffaction}
A_{eff}& = & -\hbar \sum \limits_{\nu = c,s} g_{\nu}^{-1}
\int  \dot
\theta_{\nu}(\tau)d\tau \ln(|\tau - \tau'|\Lambda_{0})\dot
\theta_{\nu}(\tau')d\tau'\nonumber\\
& + &\int d\tau V(\theta_{c}, \theta_{s})
\end{eqnarray}
where the integration in the first term is limited by the condition
$|\tau - \tau'|\Lambda_{0} \ge 1$ and the dots indicate derivatives.

The domain wall of the field theory
(\ref{cgeffaction}) can be found by solving equations $\delta
A_{eff}/ \delta \theta_{c,s} = 0$ subject to the boundary conditions
$\theta_{c}(- \infty) - \theta_{c}(\infty) = \pm q$ and
$\theta_{s}(-\infty) - \theta_{s}(\infty) = \pm 1$. In the limit of
large barriers the solution will have a form of a steep rounded step
which can be approximated by
\begin{equation}
\label{dwsolution}
\dot \theta_{c}(\tau) = \pm q \delta(\tau -
\tau_{0}),~~~~\theta_{s}(\tau) = \pm \delta(\tau - \tau_{0})
\end{equation}
where $\tau_{0}$ is arbitrary position of the domain wall.  The
combination of Eq.(\ref{cgeffaction}) and (\ref{dwsolution}) implies
that in the large-barrier limit the physics is dominated by
thermally excited logarithmically interacting domain walls whose sequence is
constrained by the zigzag locations of classical ground states
discussed earlier.

\subsection{The ``entropy'' argument of Landau, Thouless and
Kosterlitz and Thouless}

We begin with a semi-quantitative argument similar to the
``entropy'' arguments of Landau \cite{entropy}, Thouless \cite{Thouless}, and
Kosterlitz and Thouless \cite{KT}.  The energy of a single domain wall
diverges with the system size $L$, as indicated by
Eqs.(\ref{cgeffaction}) and (\ref{dwsolution}):  $E = \hbar
(q^{2}g_{c}^{-1} + g_{s}^{-1})\ln(\Lambda_{0}L)$.  The entropy is
given by the logarithm of the number of ways to put the domain wall on
the line: $S =\ln(\Lambda_{0}L)$.  Therefore we find
for the change of the free energy $F = E - TS$ for the introduction of
a single domain wall
\begin{equation}
\label{fenergy}
F = \left (\hbar({\frac {q^{2}}{g_{c}}} + {\frac {1}{g_{s}}}) -
T\right ) \ln(\Lambda_{0}L)
\end{equation}
The free energy changes sign at the temperature
\begin{equation}
\label{tc}
T_{c} = \hbar({\frac {q^{2}}{g_{c}}} + {\frac {1}{g_{s}}})
\end{equation}
This suggests that above $T_{c}$ the ground-state is unstable against
spontaneous creation of domain walls.  Below $T_{c}$ there are only
bound pairs of domain walls while above $T_{c}$ there are free domain
walls.

To see the implications of these conclusions for the problem of a nearly-closed Coulomb-blockaded quantum dot we note that in the mapping
between quantum field-theoretical models and those of classical statistical mechanics \cite{Kogut}
zero-point motion corresponds to thermal fluctuations, i. e. the role
of Planck's constant $\hbar$ is played by the temperature $T$.  Therefore
provided
\begin{equation}
\label{criterion}
1 - {\frac {1}{g_{s}}} - {\frac {q^{2}}{g_{c}}} < 0
\end{equation}
weak tunneling will not smear the Coulomb staircase.  If this is the
case, then there is non-zero lower limit to the magnitude of the population
jump $\Delta \theta_{c}$ (given by $q$) at half-integer dimensionless gate voltage $N$:
\begin{equation}
\label{jump}
(\Delta \theta_{c})_{min} = (g_{c}(1 - g_{s}^{-1}))^{1/2}
\end{equation}
This parallels the conclusion that there is non-zero lower limit to the
magnetization of the one-dimensional classical Ising model with
$1/\tau^{2}$ interactions \cite{Thouless}.  A conclusion qualitatively
similar to (\ref{jump}) has been found earlier for the spinless
version of the Coulomb blockade problem \cite{KKQ}.  One may argue
that a formal transition from spinfull to spinless electrons would
consist in imposing $SU(2)$ spin symmetry, $g_{s} = 2$,  followed by
the replacement \cite{RMP} $g_{c} = 2g$.  Then Eq.(\ref{jump}) will
exactly reproduce the spinless result \cite{KKQ}.

In the presence of $SU(2)$ spin symmetry, $g_{s} = 2$, the criterion
(\ref{criterion}) becomes $g_{c} < 2q^{2}$ which for nearly closed
dot, $q = 1$, further simplifies to $g_{c} < 2$.  The latter condition is
satisfied in the presence of short-range repulsion between the
electrons \cite{RMP}.  We conclude that repulsive interactions between the
electrons aid the Coulomb staircase to survive disordering effect of the
zero-point motion.  This meets the physical expectation
as repulsions decrease quantum fluctuations in the charge channel,
thus making the system more classical.

It is curious that non-interacting electrons, $g_{c,s} = 2$, and
a nearly-closed dot, $q = 1$, correspond to the sign of
equality in Eq.(\ref{criterion}) - the ``entropy'' argument predicts
neutral stability.  This is the first indication that the free-electron case
has to be treated with special care.  Here the analysis has to be conducted
along two directions:

First, renormalization-group analysis of Section IVB improves on the ``entropy'' argument, thus allowing us to
make definite conclusions about the case of non-interacting electrons.
Second, one should try to go beyond the two-state approximation which is
in the heart of the zigzag nature of the classical ground states.
This more complicated problem will be left for future study.

In the case of inequality opposite to (\ref{criterion}), even weak
tunneling destroys the Coulomb blockade.

\subsection{Renormalization-group treatment}

One can improve on the ``entropy'' argument via a
renormalization-group analysis of the action (\ref{cgeffaction})
perturbative in dimensionless tunneling amplitude $\Delta$
(measured in units of $\hbar \Lambda_{0}$) between
nearest ground states.  Another pertinent parameter is the distance
$q$ between the nearest wells of the potential $V(\theta_{c}, \theta_{s})$
in the $\theta_{c}$ direction.  Corresponding lowest order
renormalization-group flow equations are implicit in the work of Kane
and Fisher \cite{KF}:
\begin{equation}
\label{KF1}
{\frac {d\Delta}{dl}} = (1 - {\frac {1}{g_{s}}} - {\frac {q^{2}}{g_{c}}})\Delta
\end{equation}
\begin{equation}
\label{KF2}
{\frac {d q^{2}}{dl}} = - 8q^{2} \Delta^{2}
\end{equation}
If we set \cite{note3} $g_{c} = g_{s} = 4g$,
then Eqs.(\ref{KF1}) and (\ref{KF2}) reduce to those of Kane and
Fisher \cite{KF} describing the physics of resonant tunneling of
interacting spinless electrons onto and off a dot formed by two
barriers in the presence of a gate which couples to the
electrons between the barriers.

On the other hand, setting $g_{s} = 2$ and $g_{c} = 2g$ nearly
reproduces renormalization-group equations describing spinless version
of the problem of nearly-closed Coulomb-blockaded quantum dot
\cite{KKQ};  the physics encoded in Eqs.(\ref{KF1}) and (\ref{KF2})
parallels in many ways that of the spinless case.

Comparing Eqs.(\ref{criterion}) and (\ref{KF1}) we conclude that the
``entropy'' argument of Section IVA is equivalent to first-order
renormalization-group calculation establishing the range of
stability of the zero-tunneling fixed line.

Eqs.(\ref{KF1}) and (\ref{KF2}) have to be solved subject to the
``initial'' conditions, $\Delta(l = 0) = \Delta_{0}$ and $q(l = 0) =
q_{0}$;  the values of $\Delta_{0}$ and $q_{0}$ can be computed from the form of the action
(\ref{cgeffaction}).
\begin{figure}
\includegraphics[
  width=1.0\columnwidth,
  keepaspectratio]{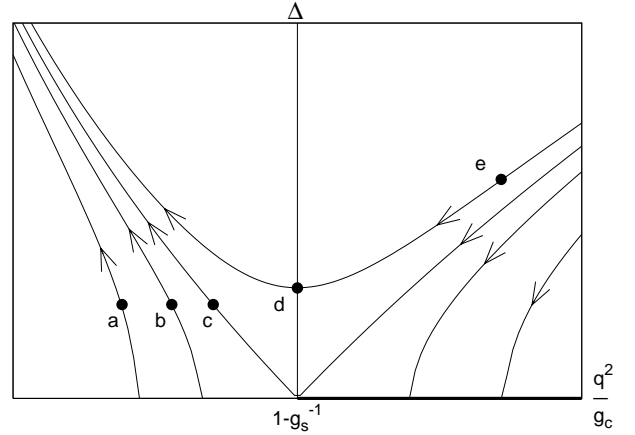}
\caption{The flow diagram corresponding to Eqs. (\ref{KF1}) and
(\ref{KF2}).  The stable part of the $\Delta = 0$ fixed line is shown
in bold, and the arrows indicate the direction of the flow.  The
models whose loci are given by the ``initial'' conditions ``a''-``e'' all exhibit smeared Coulomb staircase with
characteristic flow parameter $l_{c}$, 
given by Eqs.(\ref{lc1})-(\ref{lc5}), respectively.  Specifically, 
non-interacting electrons, $g_{c.s} = 2$, entering a nearly-closed dot,
$q_{0} = 1$,
are represented by the ``initial'' condition 'd''.}
\end{figure}

The flow diagram corresponding to Eqs.(\ref{KF1}) and (\ref{KF2}) is
sketched in Fig.2.  As typical of the Kosterlitz-Thouless phase
transition \cite{KT}, the flows can be divided into two main regions -
presently they correspond to the loci of the models where the Coulomb blockade
either survives disordering effect of zero-point motion or ceases to
exist.

In the former case the systems satisfying
$1 - g_{s}^{-1} - q_{0}^{2}/g_{c} \le 0$ (compare with
Eq.(\ref{criterion})) and $4\Delta_{0}^{2} \le q_{0}^{2}/g_{c}
+ g_{s}^{-1} - 1 - (1 - g_{s}^{-1})\ln(q_{0}^{2}/g_{c}(1 -
g_{s}^{-1}))$, i. e. those to the right of the separatrix $4\Delta^{2} = q^{2}/g_{c}
+ g_{s}^{-1} - 1 - (1 - g_{s}^{-1})\ln(q^{2}/g_{c}(1 -
g_{s}^{-1}))$ are carried by the flow to the stable fixed line, $\Delta
= 0$, $1 - {g_{s}}^{-1} - q^{2}/g_{c} \le 0$.  This indicates that
tunneling is irrelevant in macroscopic limit, and the discontinuity of
the dot population, $\Delta \theta_{c}$ (given by $q$), at half-integer $N$ satisfies the inequality $(\Delta \theta_{c})_{min} \ge
(g_{c}(1 - g_{s}^{-1}))^{1/2}$.  The equality is
reached at the Kosterlitz-Thouless phase transition point, $4\Delta_{0c}^{2} = q_{0}^{2}/g_{c}
+ g_{s}^{-1} - 1 - (1 - g_{s}^{-1})\ln(q_{0}^{2}/g_{c}(1 -
g_{s}^{-1}))$, where the population jump (\ref{jump}) depends
\textit{only} on interparticle interactions.  This universal
relationship which is a general feature of a Kosterlitz-Thouless
transition \cite{KT} is analogous to that between the magnetization
jump, the  amplitude of the $1/\tau^{2}$ interaction and the phase
transition temperature of the Ising model \cite{Thouless},\cite{Ising}.

The systems to the left of the separatrix $4\Delta^{2} = q^{2}/g_{c}
+ g_{s}^{-1} - 1 - (1 - g_{s}^{-1})\ln(q^{2}/g_{c}(1 -
g_{s}^{-1}))$ are taken by the flow outside of the range of
applicability of Eqs.(\ref{KF1}) and (\ref{KF2}), the effective
tunneling amplitude, $\Delta$, is seen to increase while $q$, the distance between
the minima of the potential $V(\theta_{c}, \theta_{s})$ in the
$\theta_{c}$ direction, tends to zero.  We are thus in the regime of
smeared Coulomb staircase with continuous dependence of the dot
population on the gate voltage.  Although the Kosterlitz-Thouless
transition is continuous, in going from sharp to
smeared Coulomb staircase, the shape of the staircase changes
\textit{discontinuously}.

In the classical limit of closed dot, $\Delta_{0} = 0$, the distance
between the minima of the potential $V(\theta_{c}, \theta_{s})$ in the
$\theta_{c}$ direction is unity, $q_{0} = 1$.   Then $SU(2)$ symmetric
models with electron-electron repulsion,
i. e. those satisfying $g_{s} = 2$, $g_{c} < 2$, belong to the
stable part of the fixed line, $\Delta
= 0$, $1 - {g_{s}}^{-1} - q^{2}/g_{c} \le 0$.  Allowing weak
tunneling, i. e. sufficiently small $\Delta_{0}$, only renormalizes
the population jump at half-integer gate voltage $N$ downward so that
$0 < \Delta \theta_{c} < 1$.  This supports the conclusion of
Section IVA that weak tunneling does not destroy the Coulomb
blockade in the presence of interparticle repulsions.  However at
sufficiently large tunneling amplitude $\Delta_{0c}$ the
Kosterlitz-Thouless transition takes place, and for $\Delta_{0} >
\Delta_{0c}$ the Coulomb staircase ceases to exist.

In this same classical limit, $\Delta_{0} = 0$, $q_{0} = 1$, but with the electrons
in the channel satisfying $1 - {g_{s}}^{-1} - g_{c}^{-1} > 0$,
we find ourselves on the unstable part of the $\Delta = 0$ fixed
line.  Here allowing infinitesimally small tunneling, $\Delta_{0}$,
immediately destroys the classical staircase which is consistent with
the ``entropy'' argument.

The free-electron case, $g_{c,s} = 2$, is marginal
because classically, $\Delta_{0} = 0$, $q_{0} = 1$, it corresponds to the
end point of the zero-tunneling, $\Delta = 0$, stable fixed line.  However allowing
infinitesimally small tunneling, as indicated in Fig.2, immediately moves it into the regime
where the tunneling amplitude grows under renormalization, and thus the
Coulomb blockade is destroyed.  This is in agreement with the original
result \cite{Matveev91}.

The systems where the Coulomb blockade is destroyed, i. e. those which
under renormalization transformation are taken outside the range of applicability of
perturbative, $\Delta \ll 1$, equations (\ref{KF1}) and (\ref{KF2}), can be
characterized by the characteristic flow parameter $l_{c}$ for which $\Delta
(l_{c}) \simeq 1$.  The corresponding frequency scale,
$\Lambda_{c} = \Lambda_{0} e^{-l_{c}}$ (see Eq.(\ref{lambdaofl}))
marks the onset of strong charge (electron number) fluctuations on
large time scales:

At times shorter than $\Lambda_{c}^{-1}$, the tunneling remains
weak, $\Delta(l)  \ll 1$, and the dot population is
frozen at an integer value.  On the contrary, at times larger than
$\Lambda_{c}^{-1}$, the tunneling is strong and the population of the
dot fluctuates freely (around a half-integer value) thus making the
particle discreteness irrelevant. The ``spin'' counterpart of
the parameter $l_{c}$ (see Eq.(\ref{lc})) plays a related role in
marking the onset of freezeout of spin fluctuations pertinent to
understanding the physics of nearly-open dot.

The estimate of $l_{c}$ cannot be expected to be more than qualitative as the perturbative
equations (\ref{KF1}) and (\ref{KF2}) are at most only indicative of its
presence:

If classically the model in question belongs to the unstable part of
the zero tunneling fixed line, then the flow starts
out vertically away from the $\Delta = 0$ axis.  If additionally the
beginning point is not too close to the end point
$\Delta = 0$, $1 - g_{s}^{-1} - q^{2}g_{c}^{-1} = 0$ (initial
condition ``a'' in Fig.2), then to lowest order in $\Delta_{0}$, the
renormalization of $q$ given by Eq.(\ref{KF2}) can be neglected.  With
this in mind the characteristic flow parameter $l_{c}$ can be found to be
\begin{equation}
\label{lc1}
l_{c} = {\frac {1}{1 - g_{s}^{-1} - g_{c}^{-1}}}\ln\left ({\frac
{1}{\Delta_{0}}}\right )
\end{equation}

If, on the other hand, the beginning point of the flow is close to the end point
$\Delta = 0$, $1 - g_{s}^{-1} - q^{2}g_{c}^{-1} = 0$ (initial
condition ``b'' in Fig.2), then renormalization of $q$, given by
Eq.(\ref{KF2}) is no longer negligible;  instead we find
\begin{equation}
\label{lc2}
l_{c} = {\frac {1}{1 - g_{s}^{-1} - g_{c}^{-1}}}\ln\left ({\frac {(1 -
g_{s}^{-1} - g_{c}^{-1})}{\Delta_{0}(1 - g_{s}^{-1})^{1/2}}}\right )
\end{equation}
We note that the estimates (\ref{lc1}) and (\ref{lc2}) are only
logarithmically sensitive to the uncertainty, $\Delta(l_{c}) \simeq
1$,  built into the definition of the characteristic flow parameter
$l_{c}$.

If the beginning point of the flow belongs to the outgoing leftward
separatrix in Fig.2 (initial condition ``c''), then to lowest order in
$\Delta_{0}$ the characteristic flow parameter $l_{c}$ is given by
\begin{equation}
\label{lc3}
l_{c} = {\frac {1}{2\Delta_{0}(2(1 - g_{s}^{-1}))^{1/2}}}
\end{equation}
while if the initial point is at the minimum of the flow trajectory,
$1 - g_{s}^{-1} - q^{2}g_{c}^{-1} = 0$ (initial condition ``d'' in
Fig.2), then we find
\begin{equation}
\label{lc4}
l_{c} = {\frac {\pi}{4\Delta_{0}(2(1 - g_{s}^{-1}))^{1/2}}}
\end{equation}
The models located between the initial conditions ``c'' and ``d'' in
Fig. 2 have characteristic flow parameter $l_{c}$ behaving as  $\Delta_{0}^{-1}(1
- g_{s}^{-1})^{-1/2}$ and interpolating between the estimates
(\ref{lc3}) and (\ref{lc4}).  We note that the systems represented by
the initial condition ``d'' in Fig.2 include non-interacting electrons,
$g_{c,s} = 2$, for which the characteristic flow parameter is
$l_{c} = \pi/4\Delta_{0}$.

If the model in question lies close to the point of
Kosterlitz-Thouless phase transition separating the regimes of sharp
and smeared Coulomb staircase, i. e. slightly higher than the incoming
rightmost separatrix (initial condition ``e'' in Fig. 2), then the
effective tunneling rate $\Delta(l)$ initially decreases with $l$ followed
by an increase at larger $l$.  As a result the characteristic
flow parameter
\begin{equation}
\label{lc5}
l_{c} = {\frac {\pi}{4\sqrt {(1 - g_{s}^{-1})\Delta_{0c}(\Delta_{0} -
\Delta_{0c}))}}}
\end{equation}
is dominated by the ``time'' spent in the
vicinity of the end point $\Delta = 0$, $1 - g_{s}^{-1} -
q^{2}g_{c}^{-1} = 0$.  Here $\Delta_{0c}$ is the tunneling amplitude at the
phase transition and it is assumed that $\Delta_{0} - \Delta_{0c} \ll
\Delta_{0c}$.  As expected, the corresponding scale in imaginary time
direction $L_{c} = \Lambda_{0}^{-1} e^{l_{c}}$ has a functional form
characteristic of the behavior of the correlation length in the
vicinity of the Kosterlitz-Thouless phase transition \cite{KT}.

The curious feature of the leading order estimates
(\ref{lc3})-(\ref{lc5}) is that they are insensitive to the
uncertainty $\Delta(l_{c}) \simeq 1$ built into the definition of the
characteristic flow parameter $l_{c}$.

\subsection{Vicinity of half-integer dot population:  heuristic argument}

One of the most intriguing statements made by Matveev is the
argument that (for non-interacting electrons) the weak-tunneling
results are implied by those found for a nearly-open dot
\cite{Matveev95}.

Renormalization-group analysis of Section IVB enables us to see how
and why the types of non-analytic behavior of the
ground-state properties found for a nearly-open dot (Section III) may be
applicable well beyond that regime:

This is expected to be the case for every system which under renormalization
transformation flows outside the range of applicability of perturbative
equations (\ref{KF1}) and (\ref{KF2}), i. e. into the region of large
tunneling amplitude $\Delta$ and vanishing distance $q$ between the minima
of the potential $V(\theta_{c},\theta_{s})$, Eq.(\ref{effaction}).
Since this is the regime of a nearly-open dot, it seems plausible that
no new physics intervenes for intermediate tunneling.   As far the
the weak-tunneling regime goes, the description of Section III must
take over at frequencies smaller than $\Lambda_{c} = \Lambda_{0} e^{-l_{c}}$.

As a test of this statement we now show how the
weak-tunneling results can be guessed from their strong-tunneling
counterparts.  First we remind the
reader that as charge fluctuations renormalize the distance $q$ between
the zigzag minima of the effective potential
$V(\theta_{c},\theta_{s})$ in the $\theta_{c}$ direction, the step of
the zigzag in the $\theta_{s}$ direction remains unaffected.  The latter
periodicity becomes ``invisible'' for $q = 0$ and half-integer
dimensionless gate voltage because the effective potential ceases to
be dependent on the spin field $\theta_{s}$.

The periodicity with respect to the spin field
$\theta_{s}$ becomes ``visible'' again if the dimensionless gate voltage
$N$ is tuned away from half-integer value.  For small deviation
the effective
potential will have be proportional to $\delta N$, and the overall
energy scale of the potential will be given by the charging energy
$E_{C}$.  These arguments imply the estimate
\begin{equation}
\label{estimate}
V(\theta_{s}) \simeq E_{C} \delta N \cos\pi \theta_{s}
\end{equation}
where in displaying the $\theta_{s}$ periodicity we restricted
ourselves to the leading Fourier harmonic and set zero of $V$ at
$\delta N = 0$.   The potential of exactly the same functional form
enters the effective action describing the strong-tunneling dynamics
of the spin degrees of freedom (\ref{effspinaction}), and the vicinity
of a half-integer dimensionless gate voltage, $\delta N \ll 1$,
guarantees the applicability of the results of Section III.  Keeping in mind that the role of
the frequency cutoff $\Lambda_{0}$ will be played by $\Lambda_{c} =
\Lambda_{0} e^{-l_{c}}$, we find the correspondence
\begin{equation}
\label{correspondence}
v_{0} \rightarrow - {\frac {E_{C}\delta N}{\hbar \Lambda_{c}}} \simeq -
\delta N e^{l_{c}}
\end{equation}
where we assumed a fixed ratio of the charging energy $E_{C}$ to the
original cutoff energy $\hbar \Lambda_{0}$.  Therefore the ground-state properties of nearly-closed quantum dot in
the vicinity of half-integer dimensionless gate voltage can be
extracted from the strong-tunneling results (Section III) provided the
substitution (\ref{correspondence}) is made and the frequency cutoff
$\Lambda_{0}$ is replaced with $\Lambda_{c} = \Lambda_{0} e^{-l_{c}}$.

For example, making these changes in Eq.(\ref{estimateg_s=2}) we
find that in the presence of $SU(2)$ spin symmetry, $g_{s} = 2$, the
non-analytic correction
to the ground-state energy in the weak-tunneling regime has the
form
\begin{equation}
\label{wtestimate}
\delta E \simeq  - E_{C} e^{l_{c}} \delta N^{2}
\ln\left ({\frac {1}{\delta N^{2}}}\right )
\end{equation}
For non-interacting electrons, when $l_{c} = \pi/4\Delta_{0}$
(see Eq.(\ref{lc4})), and to leading order in the tunneling amplitude, this
reproduces the functional form derived earlier \cite{Matveev91} via
the analogy to the two-channel Kondo problem and the exact solution of the
latter.

Our analysis implies that non-analytic behavior exhibited in
Eq.(\ref{wtestimate}) is also characteristic of any system with $SU(2)$
spin symmetry having not very strong interparticle repulsions and
sufficiently large tunneling amplitude.  Specifically, in the
vicinity of the critical tunneling amplitude $\Delta_{0c}$ which
corresponds to the Kosterlitz-Thouless transition into the regime of
sharp Coulomb staircase, the correction to the ground-state energy is
given by the combination of Eqs.(\ref{lc5}) and (\ref{wtestimate}).

\section{Conclusions}

In this paper we analyzed the zero-temperature dynamics of electrons
hopping onto and off a large quantum dot connected to a reservoir via
a one-dimensional
channel and interparticle interactions inside the channel as well as
Coulomb interactions among the electrons on the dot were taken into
account.  The main analytical tool
applied to understand both the limits of strong and weak tunneling was
the renormalization-group method.

For strong tunneling (nearly-open dot) the charge discreteness is
irrelevant and the physics is dominated by
fluctuating spin degrees of freedom which are responsible for
non-analytic behavior of the ground-state properties at half-integer
dot population.  Specifically, the previously found logarithmic
divergence of the dot capacitance \cite{Matveev95} is demonstrated
to be a consequence of the assumption that electrons in the
are non-interacting.  We show that the same conclusion
holds for physical systems possessing $SU(2)$ spin symmetry.  More
generally, however, the logarithmic singularity is replaced by a power
law non-analyticity whose exponent is determined by the degree of spin
fluctuations.

For strong tunneling (nearly-closed dot) and half-integer dot
population the spin degrees of freedom play a secondary role and the
physics is dominated by zero-point hops of the electrons onto and off
the dot.  If these fluctuations are sufficiently strong,
the Coulomb blockade is destroyed by arbitrarily weak tunneling.
Non-interacting electrons belong to this regime which is in agreement
with previous studies \cite{Matveev91}.  If the charge fluctuations
are sufficiently weak, then the Coulomb blockade survives
provided the tunneling is not too strong.  The regimes of sharp and
smeared Coulomb staircases meet at a phase transition of the
Kosterlitz-Thouless kind.

Renormalization-group analysis of the weak-tunneling regime also makes
a case in favor of the argument (first given for
non-interacting electrons \cite{Matveev95}) that non-analytic behavior
of the ground-state properties at half-integer dot population found in
the limit of nearly-open dot can extend well beyond the regime of
strong tunneling.

Finally, renormalization-group treatment of weak tunneling rests
on the two-state approximation in which all the population states
of the quantum dot separated from the two degenerate minima by energy gaps are
ignored.  Although physically plausible, similar to the spinless
version of the problem \cite{KKQ}, its applicability to the case
of non-interacting electrons requires further justification.  Indeed, in
the classical limit of closed dot, $\Delta_{0} = 0$, $q_{0} = 1$, the locus of
non-interacting electrons, $g_{c,s} = 2$, coincides with the end point
of the zero-tunneling fixed line in Fig.2.  Allowing small finite
tunneling $\Delta_{0}$ brings this model into the unstable
regime characterized by the initial condition ``d''.  If we now assume
that the ignored metastable states are populated with small finite
probability, then the actual electron number fluctuations will be suppressed
stronger than predicted by the two-state approximation.

Heuristically, the effect of weaker zero-point motion can still be
seen within the two-state analysis - the initial point of the
flow will have to be located somewhat to the right of the initial
condition ``d'' in Fig.2.  If the initial condition lies to the right
of the incoming separatrix of the flow,  then the Coulomb blockade can survive weak tunneling.   Additionally, the Coulomb staircase would be destroyed at
finite tunneling via a phase transition.
This possibility calls for an investigation which avoids the two-state
truncation altogether.  Variational analysis of this problem which
does not rely on the two-state approximation and similar to the one
used in the spinless case \cite{KKQ} will be presented elsewhere
\cite{unpublished}.

\section{ACKNOWLEDGMENTS}

This work was supported by the Chemical Sciences, Geosciences and Biosciences
Division, Office of Basic Energy Sciences, Office of Science, 
U. S. Department of Energy, and by the Thomas F. Jeffress and Kate
Miller Jeffress Memorial Trust.


\begin{thebibliography} {10}

\bibitem{Aleiner}  I. L. Aleiner, P. W. Brouwer, and L. I. Glazman, Phys. Rep.
{\bf 358}, 309 (2002), and references therein.

\bibitem{Likharev} D. V. Averin and K. K. Likharev, in 
\textit{Mesoscopic Phenomena in Solids} edited by B. L. Altshuler 
\textit{et al.} (Elsevier, Amsterdam, 1991), p. 173, and references
therein;  apparently the first observations of the effect of the
Coulomb blockade have been reported in I. Giaever and H. R. Zeller,
Phys. Rev. Lett. \textbf{20}, 1504 (1968) and J. Lambe and R. C. Jaklevic,
Phys. Rev. Lett. \textbf{22}, 1371 (1969). 

\bibitem{glaz} L. I. Glazman and K. A. Matveev,  Zh. Eksp. Teor. Fiz., 
\textbf{98}, 1834 (1990) [Sov. Phys. JETP, \textbf{71}, 1031 (1990)].
  
\bibitem{Matveev91} K. A. Matveev, Zh. Eksp. Teor. Fiz., \textbf{99}, 
1598 (1991) [Sov. Phys. JETP, \textbf{72}, 892 (1991)].

\bibitem{Flensberg} K. Flensberg, Phys. Rev. B \textbf{48}, 11156 (1993).

\bibitem{Matveev95} K. A. Matveev, Phys. Rev. B \textbf{51}, 1743 (1995).

\bibitem{KKQ} E. B. Kolomeisky, R. M. Konik, and X. Qi,  
Phys. Rev. B \textbf{66}, 075318 (2002).

\bibitem{Berman} D. Berman, N. B. Zhitenev, R. C. Ashoori, and M. Shayegan,
Phys. Rev. Lett. \textbf{82}, 161 (1999).    

\bibitem{RMP}  E. B. Kolomeisky and J. P. Straley,
Rev. Mod. Phys. \textbf{68}, 175 (1996), and references therein. 

\bibitem{Ma}  S. - K. Ma, {\it Modern Theory of Critical Phenomena}, 
(Benjamin, Reading, MA, 1980), and references therein.

\bibitem{KF}  C. L. Kane and M. P. A. Fisher, Phys. Rev. B
\textbf{46}, 15233 (1992).

\bibitem{Kogut}  J. B. Kogut, Rev. Mod. Phys. \textbf{51}, 659 (1979),
and references therein.

\bibitem{Peierls1} The classical counterpart of the action 
(\ref{effspinaction})
is implicit in R. E. Peierls, Helv. Phys. Acta \textbf{7}, Suppl. II,
81 (1934); see also R. E. Peierls, in \textit{Surprises in Theoretical
Physics} (Princeton University Press, Princeton, NJ, 1979), Sec. 4.1. 

\bibitem{LL1}  L. D. Landau and E. M. Lifshitz, \textit{Statistical
Physics}, vol.V, Part 1, third edition, revised and enlarged by
E. M. Lifshitz and L. P. Pitaevskii (Pergamon, 1980). 

\bibitem{cosint}  M. Abramowitz and I. A. Stegun, eds. 
\textit{Handbook of Mathematical Functions with Formulas, Graphs, and
Mathematical Tables} (Dover, New York, 1972, Section 5.2.

\bibitem{Nozieres}  P. Nozi{\`e}res in \textit{Solids far
from Equilibrium}, edited by C. Godr{\'e}che (Cambridge University
Press, Cambridge, 1992), p. 1, and references therein; P. Nozi{\`e}res
and F. Gallet, J. Physique, \textbf{48}, 353 (1987).

\bibitem{SG}  A. Schmid, Phys. Rev. Lett. \textbf{51}, 1506 (1983);
F. Guinea, V. Hakim, and A. Muramatsu, \textit{ibid.}, \textbf{54},
263 (1985); M. P. A. Fisher and W. Zwerger, Phys. Rev. B \textbf{32},
6190 (1985);  E. B. Kolomeisky, Pis'ma
Zh. Eksp. Teor. Fiz. \textbf{47}, 254 (1988) [JETP Lett. \textbf{47},
307 (1988); C. L. Kane and M. P. A. Fisher,
Phys. Rev. Lett. \textbf{68}, 1220 (1992);  E. B. Kolomeisky and
J. P. Straley, \textit{ibid.}, \textbf{76}, 2930 (1996).
    
\bibitem{Knops}  H. J. F. Knops and L. W. J. den Ouden, Physica,
\textbf{103}, 597 (1980). 

\bibitem{Peierls}  R. E. Peierls, \textit{Quantum Theory of Solids} (Oxford
University Press, Oxford, 1955), Sec. 5.3; \textit{More
Surprises in Theoretical Physics} (Princeton University Press,
Princeton, NJ 1991), Sec. 2.3. 

\bibitem{KS1}  E. B. Kolomeisky and J. P. Straley, Phys. Rev. B 
\textbf{53}, 12553 (1996), and references therein.

\bibitem{note2}  The parameter $K$ used by Kane and Fisher \cite{KF}
is identical to $q^{2}$ of this paper.   

\bibitem{Thouless}  D. J. Thouless, Phys. Rev. \textbf{187}, 732
(1969).

\bibitem{entropy}  Sec.163 of Landau and Lifshitz \cite{LL1}.

\bibitem{KT}  J. M. Kosterlitz and D. J. Thouless, J. Phys. C
\textbf{6}, 1181 (1973); J. M. Kosterlitz, \textit{ibid.} \textbf{7},
1046 (1974).

\bibitem{note3}  This can be arrived at by comparing Eqs.(4.2) of Kane
and Fisher \cite{KF} with Eq.(\ref{effactionnearlyopen}) of this work.

\bibitem{Ising}  P. W. Anderson and G. Yuval, J. Phys. C \textbf{4},
607 (1971);  A. J. Bray amd M. A. Moore, Phys. Rev. Lett. \textbf{49},
1545 (1982); S. Chakravarty, \textit{ibid.} \textbf{49}, 681 (1982).

\bibitem{unpublished}  E. B. Kolomeisky, unpublished.
    
\end{thebibliography}
\end{document}